# On Hierarchical Statistical Static Timing Analysis

Bing Li, Ning Chen, Manuel Schmidt, Walter Schneider, Ulf Schlichtmann
Technische Universitaet Muenchen
Arcisstrasse 21, 80333 Munich, Germany
Email: {b.li, ning.chen, manuel.schmidt, walter-karl.schneider, ulf.schlichtmann}@tum.de

*Abstract*—Statistical static timing analysis deals with the increasing variations in manufacturing processes to reduce the pessimism in the worst case timing analysis. Because of the correlation between delays of circuit components, timing model generation and hierarchical timing analysis face more challenges than in static timing analysis. In this paper, a novel method to generate timing models for combinational circuits considering variations is proposed. The resulting timing models have accurate input-output delays and are about 80% smaller than the original circuits. Additionally, an accurate hierarchical timing analysis method at design level using pre-characterized timing models is proposed. This method incorporates the correlation between modules by replacing independent random variables to improve timing accuracy. Experimental results show that the correlation between modules strongly affects the delay distribution of the hierarchical design and the proposed method has good accuracy compared with Monte Carlo simulation, but is faster by three orders of magnitude.

## I. Introduction

As the feature size of semiconductor devices scales to the deep sub-micron realm, parameter variations have stronger impact on circuit performance. These variations make traditional corner-based static timing analysis (STA) too pessimistic. Therefore, statistical static timing analysis (SSTA) is introduced to analyze circuit performance statistically. In SSTA, cell delays are modeled as functions of random variables representing process parameters with variations. Then, arrival times are propagated to compute the circuit delay. Unlike the result of STA, the circuit delay in SSTA is a distribution providing delay-yield information to designers.

Linear models are proposed in [1]–[3], where cell delays are modeled as linear functions of Gaussian random variables. This assumption simplifies the arrival time propagation algorithm at the expense of accuracy. To improve modeling and propagation accuracy, the canonical linear form in [2] is extended in [4] to handle nonlinear and non-Gaussian parameters. With the same purpose, quadratic models are proposed in [5]–[8]. It is also pointed out in [7] that their method can be in a general polynomial form with the tradeoff between runtime and accuracy. Another method to improve timing accuracy is proposed in [9], where delays are modeled as linear functions of Gaussian and non-Gaussian random variables. The latter are identified by independent component analysis.

Similar to migrating from transistor level to cell level, the hierarchical design style is adopted for more abstraction to overcome increasing design complexities. In a hierarchical design flow, a design is composed of a series of modules at different levels. In designs using IP (Intellectual Property) macros from third-party vendors as modules, the complete netlists of IP macros are not always available because of IP protection. Instead, timing models containing the same timing information are provided as replacement of the original netlists. Thereafter, timing analysis is performed at design level to compute the delay of the hierarchical circuit.

For STA, black-box and gray-box style timing models are proposed. In the black-box style, the input-output delay matrix of a module is directly used to represent its timing information. To reduce the size of the delay matrix, the assumption made in [10]–[12] that the timing of a module is mainly determined by a subset of the inputs (control signals) of the module can be applied. Contrary to using the delay matrix directly, the gray-box style method transforms the original netlist to a much smaller one by discarding structural details but maintaining the same input-output delays. In [13], [14] basic serial and parallel merges are introduced to reduce the model size. Additionally, a parallel to serial graph transformation algorithm to increase the possibility of applying the basic merge operations is introduced in [13]. In [15] a timing model extraction method based on biclique-star replacement is introduced. This method extends the parallel to serial algorithm in [13] to deal with more than two inputs and outputs in the transformation.

When variations are considered in hierarchical statistical timing analysis, most of the timing model extraction methods for STA are not valid any more. Even worse, the correlation between modules can not be contained in timing models easily. This makes design level timing analysis more challenging. In this paper, we propose a gray-box style method to extract statistical timing models for combinational circuits. Delay edges with small criticality are removed from the original timing graph to compress the timing model. Thereafter, a novel hierarchical timing analysis algorithm at design level using the proposed delay models is introduced, where the independent random variables in the timing models are replaced to incorporate the correlation between modules. With this replacement, the proposed hierarchical statistical timing analysis can produce a very accurate delay distribution curve compared to Monte Carlo simulation. To our best knowledge, this is the first paper handling hierarchical statistical timing analysis.

The rest of the paper is organized as follows. In Section II we will introduce the statistical timing method used in this paper. In Section III the requirement for generated timing models is formulated. Then, the timing model extraction method is proposed in Section IV. In Section V the hierarchical statistical timing analysis at design level is explained. Thereafter, the experimental results of the proposed methods applying to ISCAS85 benchmark circuits are shown in Section VI. Finally, we conclude our work in Section VII.

## II. STATISTICAL TIMING ANALYSIS

In the following sections we will use the concept *timing graph* to explain our algorithms. A timing graph $G$ is a weighted directed graph. A *vertex* $v_i$ in a timing graph corresponds to a pin of a cell. An *edge* $e_{ij}$ represents a delay between vertices $v_i$ and $v_j$, with the *weight* $d_{ij}$ denoting delay value. The delay $d_{ij}$ of a *path* $p_{ij}$, which is a set of consecutive connected edges between vertices $v_i$ and $v_j$, is the sum of the weights of all the edges on $p_{ij}$.

In this paper, a delay is modeled as a linear function of process parameters, like in [1]–[3]. For simplicity, we will discuss only one process parameter $p$ henceforth, written as

$$p = p_0 + p_g + p_l + p_r \quad (1)$$

where $p_0$ is the nominal value of the parameter. $p_g$ models the global variation and is shared by all delays. $p_l$ is the local variation specific to each delay and is correlated with each other. $p_r$ is an independent variable modeling the pure random effect in manufacturing processes. All $p_g$, $p_l$ and $p_r$ are assumed Gaussian and have zero mean.

Similar to [1], the die of the circuit is partitioned into $n$ grids. All the cells in the same grid share the same local variation. To represent the local variation $p_l$ in (1), a random variable $p_{l_i}$ ($i \in \{1, 2, \ldots, n\}$) is assigned to each grid. The correlation between $p_{l_i}$ and $p_{l_j}$ depends on the distance between the grids and is pre-characterized. The $n$ random variables $p_{l_i}$, written as a vector $\mathbf{p}_l$, have covariance matrix $\mathbf{C}$. Using principal component analysis (PCA) [16], $\mathbf{p}_l$ can be decomposed as

$$\mathbf{p}_l = \mathbf{A}\mathbf{x} \quad (2)$$

where $\mathbf{x} = [x_1, x_2, \ldots, x_n]^T$ is a set of independent Gaussian random variables with zero mean. The transformation matrix $\mathbf{A}$, formed by the eigenvectors of $\mathbf{C}$, is orthogonal, so that $\mathbf{A}^{-1} = \mathbf{A}^T$.

From (2) the variable $p_{l_i}$ can be written as a linear combination of $x_1, x_2, \ldots, x_n$. The coefficients of the linear combination are from the $i$th row of $\mathbf{A}$. Combining the assumption that an edge delay is a linear function of the process parameter $p$, we can write the edge delay in a general linear form

$$a_0 + a_g x_g + \sum_{i=1}^{n} a_i x_i + a_r x_r \quad (3)$$

where $x_g$ and $x_r$ are the same as $p_g$ and $p_r$ in (1) respectively. $x_i$ are independent components after applying PCA. $a_0$ is the nominal value of the delay. $a_g$, $a_i$ and $a_r$ are all coefficients with fixed values. Unlike in [2], we write $x_g$ separately in (3) because it is shared by delays in all modules in the hierarchical timing analysis.

Two computations are involved in timing analysis: sum and maximum. In this paper we use the method proposed in [2] to compute the sum and maximum of two random variables $A$ and $B$

$$A = a_0 + a_g x_g + \sum_{i=1}^{n} a_i x_i + a_r x_{r_a} \quad (4)$$

$$B = b_0 + b_g x_g + \sum_{i=1}^{n} b_i x_i + b_r x_{r_b} \quad (5)$$

The sum of $A$ and $B$ is computed by adding the corresponding coefficients of the random variables in $A$ and $B$ respectively. Then, $a_r x_{r_a} + b_r x_{r_b}$ is replaced by $c_r x_{r_c}$ so that the result of the sum is also in the form of (3). The coefficient $c_r$ is computed by matching the variances of $c_r x_{r_c}$ and $a_r x_{r_a} + b_r x_{r_b}$.

To compute the maximum of A and B, denoted as $\max\{A, B\}$, the tightness probability ($T_P$) [2] is firstly computed, which is the probability that $A$ is larger than $B$. When $A$ and $B$ are both Gaussian, $T_P$ can be computed by

$$T_P = Prob\{A \geq B\} = \Phi(\frac{a_0 - b_0}{\theta}) \quad (6)$$

where $\Phi$ is the cumulative distribution function of the standard Gaussian distribution. $\theta = \sqrt{\sigma_A^2 + \sigma_B^2 - 2Cov}$, where $\sigma_A^2$ and $\sigma_B^2$ are the variances of $A$ and $B$ respectively, and $Cov$ is the covariance between $A$ and $B$. According to [17], the mean ($\mu$) and variance ($\sigma^2$) of $\max\{A, B\}$ can be computed by

$$\mu = T_P a_0 + (1 - T_P) b_0 + \theta \phi(\frac{a_0 - b_0}{\theta}) \quad (7)$$

$$\sigma^2 = T_P(\sigma_A^2 + a_0^2) + (1 - T_P)(\sigma_B^2 + b_0^2)$$
$$+ (a_0 + b_0)\theta\phi(\frac{a_0 - b_0}{\theta}) - \mu^2 \quad (8)$$

where $\phi$ is the probability density function of the standard Gaussian distribution. In order to apply the sum and maximum computations iteratively to propagate arrival times, $\max\{A, B\}$ is approximated in the same form of (3), as

$$max\{A, B\} = m_0 + m_g x_g + \sum_{i=1}^{n} m_i x_i + m_r x_{r_m} \quad (9)$$

where $m_0$ is equal to $\mu$. $m_g$ and $m_i$ are computed by $m_g = T_P a_g + (1 - T_P) b_g$ and $m_i = T_P a_i + (1 - T_P) b_i$ respectively. $m_r$ is computed by matching the variance of the linear form (9) and $\sigma^2$ in (8).

## III. TIMING MODEL FORMULATION

Timing models are normally created by IP vendors or library groups as replacement of the original netlists of modules for timing analysis. A *timing model* is a timing graph consisting of a new set of edges and vertices but with the same inputs and outputs as the original timing graph. In order to guarantee the correct arrival time propagation at design level, a timing model must contain the necessary timing information from the original timing graph. Additionally, the timing model of a module should be as small as possible in order to accelerate the arrival time computation at design level.

An *arrival time* $a_i$ assigned to a vertex $v_i$ in a timing graph saves the maximum delay from the inputs to $v_i$. During hierarchical timing analysis, the arrival time $a_j$ at an output $v_j$ of a combinational module can be computed by

$$a_j = \max_{v_i \in I}\{\max_{p_{ij_k} \in P_{ij}}\{a_i + d_{ij_k}\}\} \qquad (10)$$

$$= \max_{v_i \in I}\{a_i + \max_{p_{ij_k} \in P_{ij}}\{d_{ij_k}\}\} \qquad (11)$$

$$= \max_{v_i \in I}\{a_i + M_{ij}\} \qquad (12)$$

where $I$ is the set of all the inputs of the module. $d_{ij_k}$ is the delay of $p_{ij_k}$, which denotes the $k$th path between the input $v_i$ and the output $v_j$. The set of all the paths between $v_i$ and $v_j$ is denoted by $P_{ij}$. $M_{ij}$ denotes the maximum path delay between $v_i$ and $v_j$.

From (11) we can conclude that the arrival time at an output of a module is determined by the arrival times at all the inputs of the module and the maximum delays from all the inputs to the output. When characterizing the timing model of a module, especially an IP block, the application context is unknown. For this reason, no assumption about the arrival times at the inputs should be made. On the contrary, the maximum input-output delays $M_{ij}$ in (12) are exclusively timing characteristics of the module.

For a module with $m$ inputs and $n$ outputs, we define the *delay matrix* as an $m \times n$ matrix with entries $M_{ij}$. From the analysis above, a pre-characterized timing model must have the same delay matrix as the original timing graph of the module to retain the correct timing information. For a module with a large number of inputs and outputs, the delay matrix may be too large to be used as a timing model. In the following, we will introduce a gray-box timing model extraction method based on timing graph reduction.

## IV. STATISTICAL TIMING MODEL EXTRACTION

In this section, we will propose a gray-box method to generate the statistical timing model for a combinational module. The two basic merge operations are introduced firstly. Then a method based on criticality is used to remove non-critical edges from the timing graph.

### A. Serial and parallel merge operations

Two basic edge merge operations which do not change the input-output delays of the module are involved in the gray-box model generation [13], [14]. The serial merge is illustrated in Fig. 1. If $n$ edges with sink vertices $v_{j_1} \ldots v_{j_n}$ leave the same vertex $v_k$ and $v_k$ has only one fanin edge with source vertex $v_i$, $v_k$ can be removed and the edges can be merged so that there are only direct edges between $v_i$ and $v_{j_1} \ldots v_{j_n}$. The weights of the new edges between $v_i$ and $v_{j_1} \ldots v_{j_n}$ are the sums of the weights $d_{ik}$ and $d_{kj_1} \ldots d_{kj_n}$, respectively. Similarly, this transformation can be applied in reverse direction. The parallel merge operation merges the edges with the same source and sink vertices, as illustrated in Fig. 2. A new edge is created to replace the parallel edges, with the weight equal to $\max_{k \in \{1 \ldots n\}}\{d_{ij_k}\}$.

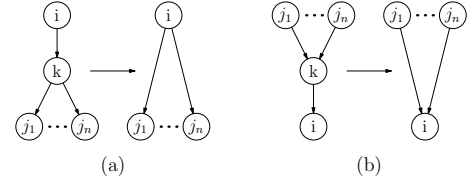

Fig. 1. Serial merge operation

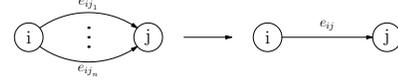

Fig. 2. Parallel merge operation

### B. Non-critical edge removal

Normally there is more than one path from an input to an output in a module. In the view of the timing analysis, only the paths with dominant delays, called *critical paths*, determine the input-output delays of the module. From this observation, the edges which are only at non-critical paths can be removed without losing modeling accuracy. Note that the definition of critical path here is different from the classical one, where the critical path dominates the paths starting from all inputs to all outputs of a circuit. In our definition, the critical path dominates all the paths starting from a specified input to a specified output.

In statistical timing analysis, all the delays are random variables. A path delay can only dominate the delay of another path with some probability. Similarly, an edge can be on the critical path only with a probability as well.

**Definition 1:** *criticality* ($c^{ij}$) *of an edge with respect to input $v_i$ and output $v_j$ is the probability that this edge is on the critical path between $v_i$ and $v_j$.*

**Definition 2:** *maximum criticality* ($c^m$) *of an edge in a module is the maximum $c^{ij}$ over all input-output pairs.*

The criticality defines the probability that an edge is on the critical path. When the criticality of an edge is very small, the edge will almost have no chance to affect the corresponding input-output delay. In order to compress the timing model, we remove the edges with criticalities with respect to all the input-output pairs less than a threshold $\delta$, i.e. $c^m < \delta$, from the timing graph. In the following, we will describe how to compute the criticality $c^{ij}$ for an edge. The computation of $c^m$ in Definition 2 is self-explanatory.

In a timing graph, there are many paths passing through an edge $e$. We denote all the paths between an input $v_i$ and output $v_j$ and passing through edge $e$ as a set $P_e^{ij}$. All the paths between $v_i$ and $v_j$ and not passing through the edge $e$ are denoted as $P_{\bar{e}}^{ij}$. The maximum of the delays of the paths in $P_e^{ij}$ and $P_{\bar{e}}^{ij}$ are written as $d_e$ and $d_{\bar{e}}$ respectively. If the edge $e$ is on the critical path, the longest path in $P_e^{ij}$ dominates the longest path in $P_{\bar{e}}^{ij}$, which means $d_e \geq d_{\bar{e}}$. This statement is also valid vice versa. Therefore, we can compute the criticality $c^{ij}$ as $Prob\{d_e \geq d_{\bar{e}}\}$. Similar to [18], we have

$$Prob\{d_e \geq d_{\bar{e}}\} = Prob\{d_e \geq d_{\bar{e}}, d_e \geq d_e\} \qquad (13)$$

$$= Prob\{d_e \geq \max\{d_e, d_{\bar{e}}\}\} \qquad (14)$$

where $\max\{d_e, d_{\overline{e}}\}$ is the maximum delay of all the paths between $v_i$ and $v_j$, and is equal to the maximum input-output delay $M_{ij}$.

Because $d_e$ is the maximum delay of the paths passing through the edge $e$, $d_e$ can be computed by (15) [18],

$$d_e = a_e + d + r_e \tag{15}$$

where $a_e$ is the maximum delay from input $v_i$ to the source vertex of $e$ and equal to the corresponding arrival time exclusively from $v_i$. $r_e$ is the maximum delay from output $v_j$ to the sink vertex of $e$ and equal to the corresponding negative required time exclusively from $v_j$, when the required time at $v_j$ is set to 0. $d$ is the delay of edge $e$. Using the propagation algorithm proposed in [19], all the maximum input-output delays $M_{ij}$ as well as the arrival times and required times for all the vertices with respect to all the input-output pairs can be computed, where the sum and maximum are computed as described in Section II. After these computations, both $M_{ij}$ and $d_e$ are in the general linear form (3), so that the probability in (14) can be computed using (6).

*C. Timing model generation*

The gray-box statistical timing model is generated by applying the algorithms introduced above sequentially, as shown in Fig. 3.

```
1. compute maximum criticality c^m for each edge
2. remove edges with c^m less than the predefined
   threshold δ
3. apply serial and parallel merge operations
   iteratively
```

Fig. 3. Gray-box timing model generation

## V. HIERARCHICAL TIMING ANALYSIS

In this section, we propose a method to propagate arrival times from primary inputs to primary outputs of a hierarchical design, using pre-characterized timing models. This method replaces the independent random variables in the timing models by a new set of random variables, so that the correlation from local variation is also taken into account.

As shown in Section II, the area of a module is partitioned into grids. The correlated local variables assigned to the grids are decomposed using PCA. Thereafter, the on-die locations of the cells inside the module are used to identify the grids they belong to so that the corresponding coefficients of the independent variables can be selected from the transformation matrix in (2).

When propagating arrival times at design level, all edge delays in a timing model are in the linear form of the independent random variables with respect to the grid partition of the die of the module. No cell layout information at design level exists in the timing models because the delay edges in the timing models are created from the original timing graphs and do not represent cell delays directly. As the result, we can not simply partition the die of the top design and run PCA to transform all delays into the linear form of the independent random variables at design level.

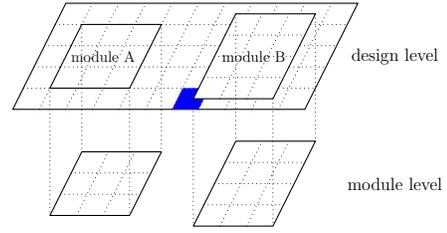

Fig. 4. Heterogeneous grids

To solve this problem, we partition the die of the top design through two steps. An example is shown in Fig. 4. At first, the die areas covered by modules are partitioned with the same grids as during timing model generation. In Fig. 4 we firstly partition the die areas covered by module A and B using the default grid size and starting from their own origins respectively, as we partitioned the die area of each module during timing model generation. Thereafter, the remaining die area which is not covered by modules is partitioned with the default grid size. All these grids together are considered as the design level grid partition. Because the origins of the modules may move freely during module layout, the design level grids may have different sizes and shapes. In Fig. 4, module A happens to have grids aligned with the grids of the second step partition, but module B has heterogeneous grids. Because the size of each grid is no larger than the default grid size, this heterogeneous partition does not lose any modeling accuracy.

For each grid at design level, a random variable is assigned to model the local variation, even though the grid is not rectangular, for example the marked grid in Fig. 4. Assuming there are totally $m$ grids after partitioning the die of the top design, the $m$ random variables are written as a vector $\mathbf{p}_l^t$, with an $m \times m$ covariance matrix $\mathbf{C}^t$. In the following, we will use module B as an example to explain the independent random variable replacement. Other modules can be processed similarly. We assume that $\mathbf{p}_l$ is the correlated random variable vector corresponding to the $n$ grids of module B during timing model generation and (2) is used to decompose $\mathbf{p}_l$. All the edge delays inside the timing model of B are linear combinations of the independent random variables $\mathbf{x}$.

Without losing generality, we assume that the random variables inside the area covered by module B at design level are indexed from 1 to $n$, denoted as $\mathbf{p}_{l,n}^t$. The $n \times n$ sub-matrix at the upper-left corner of $\mathbf{C}^t$ represents the correlation between $\mathbf{p}_{l,n}^t$. Because the correlation is determined by the distance between grids, this sub-matrix is the same as the covariance matrix $\mathbf{C}$ of the module B during timing model generation. Similar to (2), $\mathbf{p}_l^t$ can be decomposed as

$$\mathbf{p}_l^t = \mathbf{B}\mathbf{x}^t \tag{16}$$

where $\mathbf{x}^t$ are the independent random variables corresponding to $\mathbf{C}^t$. Considering only the first $n$ random variables in (16), we can write the decomposition of $\mathbf{p}_{l,n}^t$ as

$$\mathbf{p}_{l,n}^t = \mathbf{B}_n \mathbf{x}^t \tag{17}$$

where the $n \times m$ matrix $\mathbf{B}_n$ contains the first $n$ rows of the transformation matrix $\mathbf{B}$. Comparing (2) and (17), both

$\mathbf{p}_l$ and $\mathbf{p}_{l,n}^t$ are Gaussian random variable vectors with the same covariance matrix $\mathbf{C}$. $\mathbf{p}_l$ and $\mathbf{p}_{l,n}^t$ also have the same mean and variance vectors because they represent the local variation of the same process parameter. Therefore, $\mathbf{p}_l$ and $\mathbf{p}_{l,n}^t$ are equivalent. The right side of (17) shows that $\mathbf{p}_l$ can be written as linear combinations of $\mathbf{x}^t$ without affecting its covariance matrix $\mathbf{C}$, so that we have

$$\mathbf{p}_l = \mathbf{B}_n \mathbf{x}^t \quad (18)$$

From (2) and (18), we can replace the independent random variables $\mathbf{x}$ in the timing model of B by $\mathbf{x}^t$,

$$\mathbf{x} = \mathbf{A}^T \mathbf{p}_l = \mathbf{A}^T \mathbf{B}_n \mathbf{x}^t \quad (19)$$

By applying (19) to each module in the top design, the correlation between modules is modeled by sharing the new independent random set $\mathbf{x}^t$. Fig. 5 shows the complete algorithm for hierarchical timing analysis using independent variable replacement at design level.

```
1. partition the die of the top design with
   heterogeneous grids
2. decompose the correlated process parameters
   at design level using PCA
3. replace independent random variables for each
   module using (19)
4. propagate arrival times from primary inputs
   to primary outputs of the top design
```

Fig. 5. Hierarchical timing analysis

## VI. EXPERIMENTAL RESULTS

In this section, the results of the proposed method applied to the ISCAS85 benchmarks are shown. The algorithms were implemented in C++ and tested using a 2.33GHz CPU. The cells in the benchmarks were mapped to a 90nm library from an industrial partner. The standard deviations of transistor length, oxide thickness and threshold voltage were assigned to 15.7%, 5.3% and 4.4% of the nominal values respectively, in reference [20]. Load variance was assigned to 15% for our experiments. The dies of the benchmark circuits were partitioned to grids so that the number of cells in a grid is less than 100, like in [1]. The correlation of the same parameter in two neighboring grids was set to 0.92, and decreased exponentially to 0.42 when the grid distance increased to 15. All the cells with distance larger than 15 grids were assumed to have only the correlation from global variation, set to 0.42, for our experiments. The correlation between different parameters was ignored for simplicity.

### A. Results of timing model extraction

The effectiveness of the proposed timing model extraction method in Section IV relies on that there are many edges in the original timing graph with the maximum criticality $c^m$ less than the threshold $\delta$. The criticality histogram for the benchmark c7552 is shown in Fig. 6. From this histogram, we can see that the edge criticalities in the benchmark tend to 0 and 1. In our experiment, the other ISCAS85 benchmark circuits show the similar tendency too. From this observation, we can remove many edges from the timing graph without affecting the timing characteristic of the model much, if the criticality threshold $\delta$ is set to a small value, for example 0.05 in this paper.

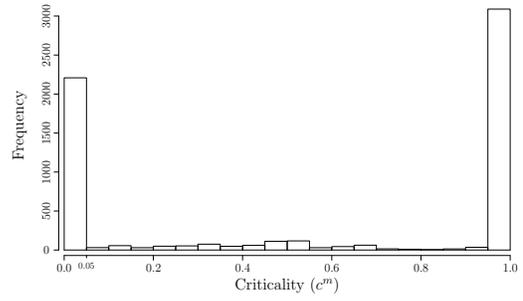

Fig. 6. The edge criticalities in c7552

The quality of a timing model is evaluated by two criteria. Firstly, we compare the numbers of edges and vertices in the timing models and in the original timing graphs. The smaller the timing model is, the faster the design level arrival time propagation can run. The second criterion is the accuracy of the timing model. According to Section III, a timing model should have the same maximum input-output delays as the original timing graph. But in statistical timing analysis, the maximum computation during arrival time propagation is only an approximation. Therefore, the generated timing models are also approximations. To verify the accuracy of the timing models, we ran Monte Carlo simulation with 10,000 iterations to calculate the means and the standard deviations of all input-output delays of the benchmarks. The comparison results are shown in Table I. $E_o$ and $V_o$ are numbers of edges and vertices in the benchmarks. $E_m$ and $V_m$ are the numbers of the edges and vertices in the timing models. $p_e$ and $p_v$ are defined as $E_m/E_o$ and $V_m/V_o$ to show the compression efficiencies respectively. From these comparisons, the proposed method can effectively generate much smaller timing models compared to the original modules. $m_{err}$ in Table I shows the maximum modeling error on mean of the input-output delays, and is defined as $\max\{|m_{model} - m_{MC}|/m_{MC}\}$, where $m_{model}$ and $m_{MC}$ are the means of the input-output delays in the generated timing models and from Monte Carlo simulation of the original netlists respectively. $v_{err}$ shows the maximum error of standard deviation and is defined similarly. These delay comparisons prove that the generated timing models are very accurate. The runtime of the timing model extraction is shown as $T$ in Table I. Because the criticalities of edges with respect to all input-output pairs should be computed in the non-critical edge removal algorithm, the runtime is roughly proportional to the product of the numbers of the inputs and the outputs of the module. This explains why the runtime does not increase monotonously with the increase of the module size.

TABLE I
RESULTS OF TIMING MODEL EXTRACTION

| Circuit | $E_o$ | $V_o$ | $E_m$ | $V_m$ | $p_e$ | $p_v$ | $m_{err}$ | $v_{err}$ | $T$(s) |
|---|---|---|---|---|---|---|---|---|---|
| c432 | 336 | 196 | 45 | 46 | 13% | 23% | 0.23% | 0.96% | 0.05 |
| c499 | 408 | 243 | 176 | 99 | 43% | 41% | 0.14% | 0.94% | 0.14 |
| c880 | 729 | 443 | 249 | 115 | 34% | 26% | 0.56% | 0.3 % | 0.21 |
| c1355 | 1064 | 587 | 143 | 99 | 13% | 17% | 0.44% | 0.26% | 0.37 |
| c1908 | 1498 | 913 | 264 | 93 | 18% | 10% | 0.82% | 1.47% | 0.36 |
| c2670 | 2076 | 1426 | 410 | 335 | 20% | 23% | 0.26% | 1.28% | 10.15 |
| c3540 | 2939 | 1719 | 440 | 141 | 15% | 8% | 0.49% | 0.72% | 0.93 |
| c5315 | 4386 | 2485 | 966 | 424 | 22% | 17% | 0.72% | 1.47% | 15.35 |
| c6288 | 4800 | 2448 | 429 | 188 | 9% | 8% | 1.03% | 1.6 % | 2.08 |
| c7552 | 6144 | 3719 | 1073 | 546 | 17% | 15% | 1.21% | 1.58% | 21.94 |
| average | | | | | 20% | 19% | 0.59% | 1.06% | |

## B. Results of hierarchical timing analysis

To test the hierarchical arrival time propagation algorithm proposed in Section V, we built an experimental hierarchical circuit by placing four c6288 modules, which are 16×16 multipliers according to [21], in two columns. The outputs of the two c6288 modules in the first column were cross-connected with the inputs of the other two modules in the second column. The four modules were placed in abutment so that the correlation was maximized. In Fig. 7 the result of the algorithm proposed in Section V is marked as *proposed method*. For comparison, the delay curve of the experimental circuit computed by propagating arrival times considering only the correlation from global variation at design level is marked as *only correlation from global variation*. Both results are compared with the result of Monte Carlo simulation with 10,000 iterations using the flattened netlist of the original circuit. From Fig. 7, we can draw the conclusion that the correlation from local variation has a remarkable effect on the circuit delay and the proposed method has good accuracy. Additionally, the proposed hierarchical analysis method using the generated timing models in this experiment is faster by three orders of magnitude than Monte Carlo simulation using the flattened netlist.

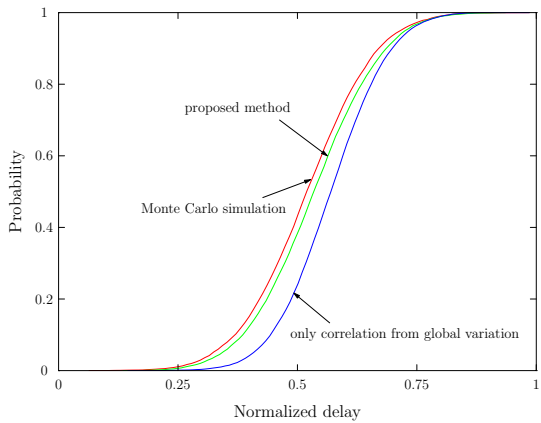

Fig. 7. Results of hierarchical timing analysis

## VII. CONCLUSION

In this paper, we proposed a method to effectively extract timing models for statistical timing analysis. Compared to the original timing graphs, the numbers of edges and vertices of the resulting models are reduced by 80% and 81% on average, respectively. Additionally, a novel independent random variable replacement algorithm was proposed to annotate the correlation from local variation to the pre-characterized timing models for hierarchical timing analysis. With such correlation, the result of the arrival time propagation at design level has very good accuracy compared with Monte Carlo simulation. Future work will incorporate the slope and load at the inputs and outputs of the modules into the timing model extraction.